\documentclass{article}
\usepackage{spconf,amsmath,graphicx}
\usepackage{cite}
\usepackage{amsthm,amsmath,amssymb,bm}
\usepackage{amsfonts}
\usepackage{booktabs}
\usepackage{url}
\usepackage{xcolor}
\usepackage{multirow}
\usepackage{makecell}
\usepackage[caption=false]{subfig}
\usepackage{float}
\usepackage[colorlinks,linkcolor=red]{hyperref}

\title{Anomalous Sound Detection using Audio Representation with Machine ID based Contrastive Learning Pretraining}

\name{Jian Guan$^{1}$, Feiyang Xiao$^1$, Youde Liu$^{2}$, Qiaoxi Zhu$^3$, Wenwu Wang$^4$
\thanks{This work was partly supported by the Natural Science Foundation of Heilongjiang Province under Grant No. YQ2020F010,  and a Newton Institutional Links Award from the British Council with Grant No. 623805725.
}
}
\address{
  $^1$Group of Intelligent Signal Processing, College of Computer Science and Technology, \\ Harbin Engineering University, Harbin, 150001, China\\
  $^2$School of Computer Science and Technology, Harbin Institute of Technology, Harbin, 150001, China\\
  $^3$Centre for Audio, Acoustics and Vibration, University of Technology Sydney, Ultimo, NSW, Australia\\
  $^4$Centre for Vision Speech and Signal Processing, University of Surrey, Guildford, GU2 7XH, UK}
\begin{document}
%
\maketitle
\begin{abstract}
Existing contrastive learning methods for anomalous sound detection refine the audio representation of each audio sample by using the contrast between the samples’ augmentations (e.g., with time or frequency masking). However, they might be biased by the augmented data, due to the lack of physical properties of machine sound,  thereby limiting the detection performance. This paper uses contrastive learning to refine audio representations for each machine ID, rather than for each audio sample. The proposed two-stage method uses contrastive learning to pretrain the audio representation model by incorporating machine ID and  a self-supervised ID classifier to fine-tune the learnt model, while enhancing the relation between audio features from the same ID. Experiments show that our method outperforms the state-of-the-art methods using contrastive learning or self-supervised classification in overall anomaly detection performance and stability on DCASE 2020 Challenge Task2 dataset.
\end{abstract}
\section{Introduction}
\label{sec:intro}
Anomalous sound detection (ASD) aims to detect the unknown anomalous sounds with only normal sounds available in training \cite{Koizumi_DCASE2020_01, suefusa2020anomalous, Giri2020a, dohi2021flow, Dohi_arXiv2022_02, GuanHEU2022, WeiHEU2022}. It has the potential in acoustic scene monitoring \cite{Giri2020a}, quality assurance \cite{chen2022learning}, and artificial intelligence-based factory automation \cite{Dohi_arXiv2022_02}. Due to the unavailability of anomalous sounds, the audio feature acquisition and representation of normal sounds is key to distinguishing the normal and unknown anomalous sounds \cite{liu2022anomalous, Giri2020b}. 

To learn the audio representation, early methods employ auto-encoder for reconstructing the input audio feature (i.e., log-Mel spectrogram) and employ the reconstruction error as the anomaly score for anomalous sound detection \cite{Koizumi_DCASE2020_01, suefusa2020anomalous}. However, these methods often ignore the use of metadata about audio files that can describe the states or properties of machines, e.g., machine ID. The operating sounds of machines with different IDs often have unique characteristics reflecting the difference between machines. In this case, the learnt audio representation may be degraded by the difference between machines of different IDs under the same machine type, which can limit the detection performance \cite{Giri2020a, liu2022anomalous}. 

As a solution, the self-supervised classification methods \cite{Giri2020a, dohi2021flow} employ machine IDs accompanying the machine sound as labels to improve the learning of audio features from different IDs, which may offer better performance \cite{Giri2020a}. However, these methods do not effectively enhance the inter-class relation between audio features with the same machine ID. As a result, the learnt features may not be sufficiently fine-grained for anomalous sound detection. These methods may perform differently even for machines of the same type, leading to instability for anomaly detection \cite{dohi2021flow, liu2022anomalous}. In our recent work, we have further improved the performance and stability of the self-supervised classification method by introducing the spectral-temporal feature, i.e., STgram \cite{liu2022anomalous}. However, the inter-class relation between the learnt features from the same ID is rarely considered. Thus, the normal and anomalous sounds from the same ID cannot be well distinguished. The learnt feature still has the potential to be further improved. Recent studies in image representation learning indicate that contrastive learning may perform better for feature learning than self-supervised classification methods \cite{chen2020simple, NEURIPS2020_d89a66c7}, and could help enhance the inter-class relation between samples \cite{NEURIPS2020_d89a66c7}.
\begin{figure*}[t]
    \centering
        \subfloat[\centering{Training procedure of the proposed CLP-SCF method.}]{
    	    \label{subfig:training_phase}
    	    \includegraphics[width=0.88\columnwidth]{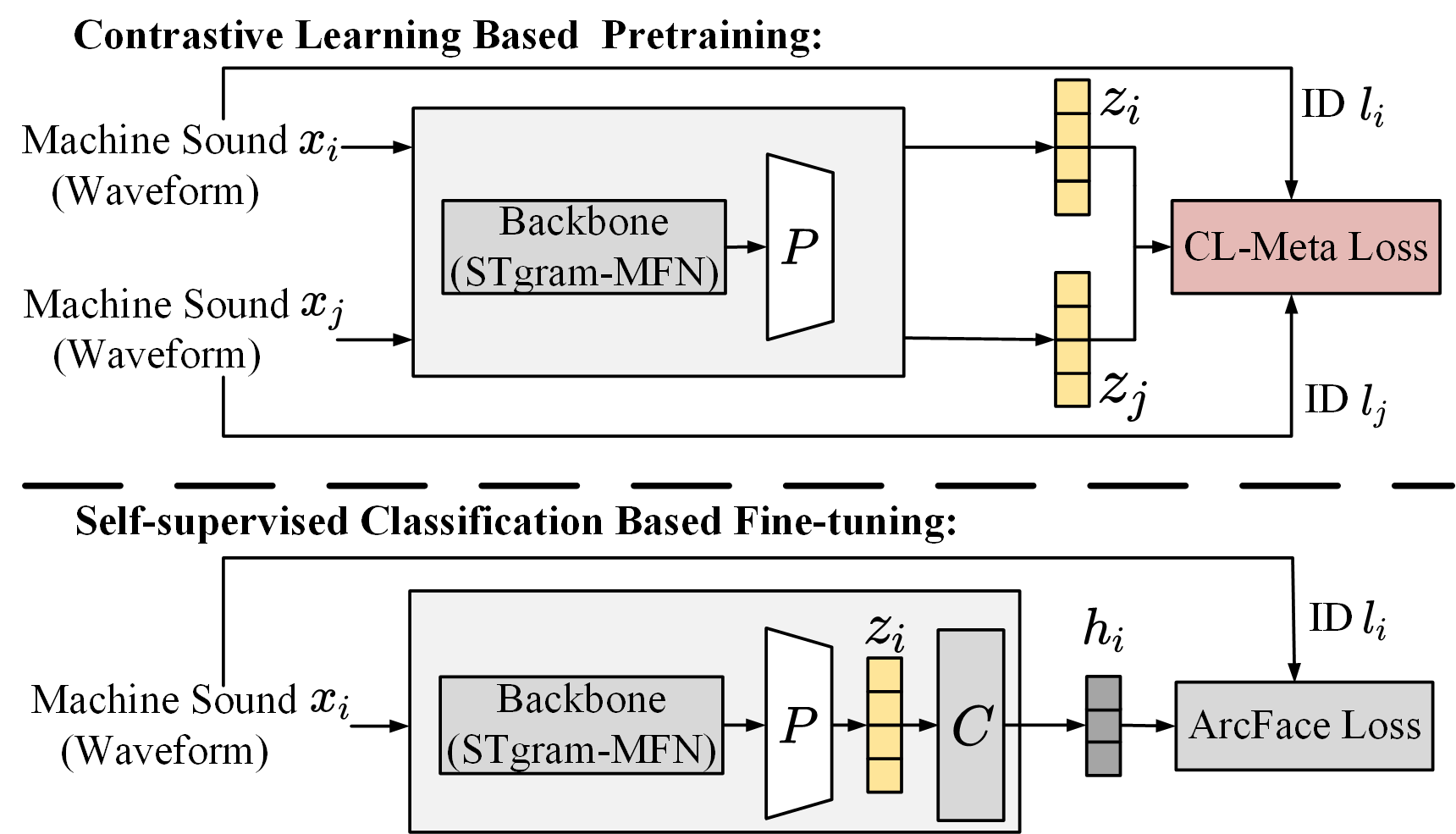}
    	    }
    	\vspace{-1.5mm}
    	\quad%
    	\subfloat[\centering{Different model structures in two stages.}]{
    	    \label{subfig:model_structure}
    	    \includegraphics[width=0.75\columnwidth]{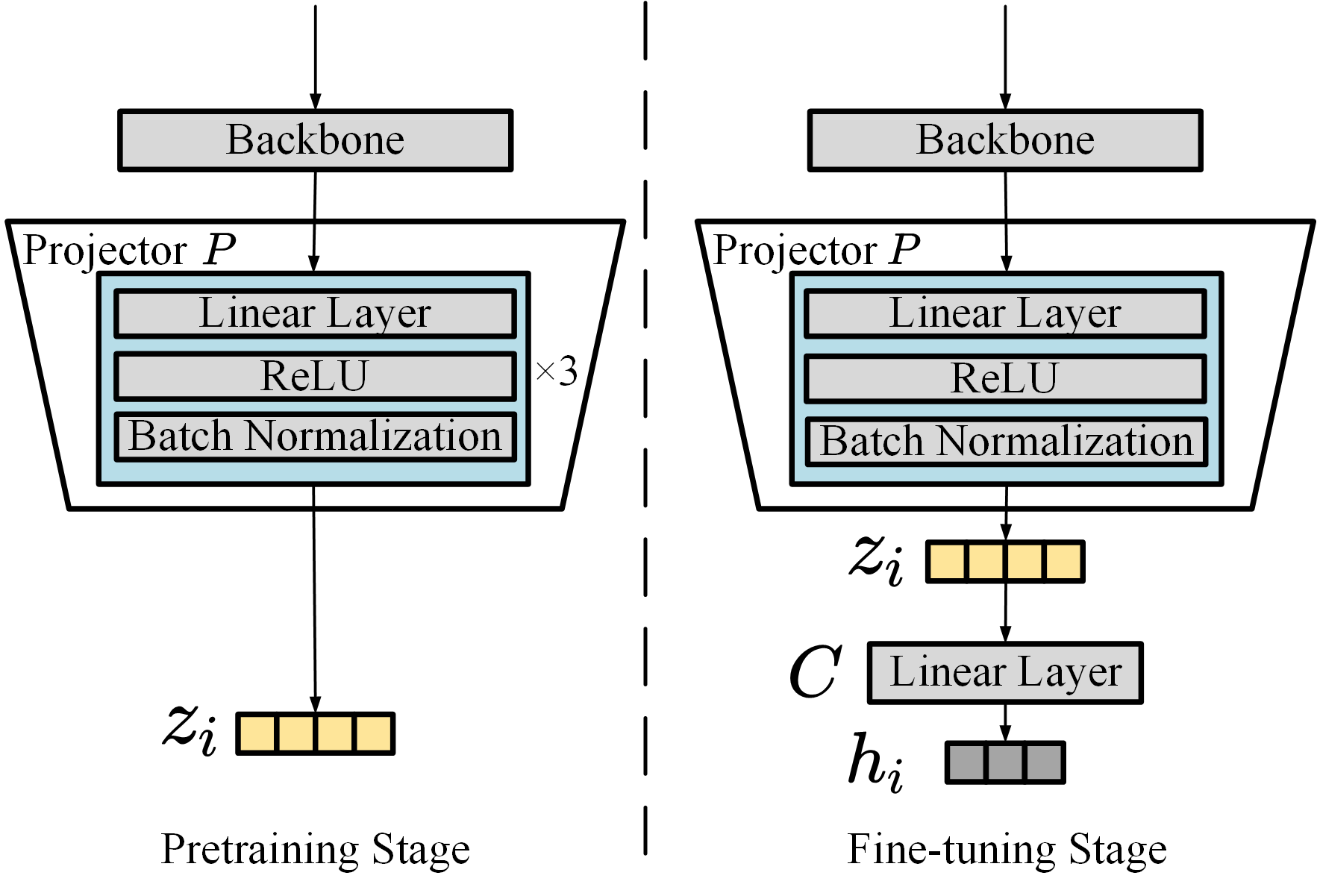}
    	    }
    	\vspace{-1.5mm}
        \caption{Framework of the proposed CLP-SCF method, where the training procedure includes two phases: contrastive learning based pretraining and self-supervised classification based fine-tuning. Different model structures are adopted in these two stages. In the pretraining stage, the model consists of a backbone module (i.e., STgram-MFN \cite{liu2022anomalous}) for audio feature extraction and a projector module $P$ for audio feature embedding. 
        In the fine-tuning stage, a self-supervised classifier $C$ is used in addition to the backbone and projector modules for fine-tuning the model. }
    \label{fig:CLPMetaSystem}
    \vspace{-5mm}
\end{figure*}

Following the success of contrastive learning in image representation, e.g., SimCLR \cite{chen2020simple}, contrastive learning has also been introduced for audio representation in recent ASD studies \cite{hojjati2022self, Cai2021}, where each audio signal is represented with audio embeddings using data augmentation (e.g., time masking, frequency masking). More specifically, the embeddings from the same audio signal are moved closer together, while the embeddings from different machine sounds are moved away from each other via contrastive processing, when learning the audio feature representation of the normal sounds. However, in practice, the embeddings of these audio signals may still be far away from each other. In other words, these methods can only learn general representation of normal machine sounds, which may not be fine-grained enough to distinguish the anomalous sounds from the normal sounds.

In this paper, we introduce contrastive learning to exploit the latent relation between the machine sound and its corresponding metadata (machine ID), and present a two-stage training method for representation learning of audio features in anomalous sound detection by combining contrastive learning based pretraining and self-supervised classification based fine-tuning (CLP-SCF). In our method, a backbone module (i.e., STgram-MFN \cite{liu2022anomalous}) is employed for audio feature extraction, and a multi-layer perception (MLP) is adopted to map the extracted audio feature to the audio embedding for contrastive learning. In the pretraining stage, a metadata-based contrastive learning (CL-Meta) loss is introduced for fine-grained feature learning, which not only increases the intra-class difference between audio features from different IDs by pushing their audio embeddings away from each other, but also enhances the inter-class relation between audio features from the same ID by clustering their audio embeddings together. In the fine-tuning stage, a self-supervised ID classifier is adopted to fine-tune our model by distinguishing the audio features of different IDs to further enhance the distinguishing ability of the learnt audio representation. Experiments conducted on DCASE 2020 dataset \cite{Koizumi_DCASE2020_01} show that the proposed method outperforms the state-of-the-art methods in both detection performance and stability.

\section{Proposed Method}
\label{sec:2}
This section presents the proposed two-stage CLP-SCF method in detail. The overall framework and the training procedure are shown in Fig.~\ref{fig:CLPMetaSystem}. In our method, a novel metadata-based contrastive learning (CL-Meta) loss is introduced for audio feature pretraining, and a self-supervised classifier is then adopted for fine-tuning the model to learn improved audio representation. Our model includes a backbone module (i.e., STgram-MFN \cite{liu2022anomalous}) for audio feature extraction and an MLP projector module to obtain audio embeddings, with different structures in two stages as shown in Fig. \ref{fig:CLPMetaSystem}(b).

\subsection{Contrastive Learning Based Pretraining}
In the pretraining stage, a novel metadata-based contrastive learning (CL-Meta) loss is introduced for audio feature pretraining. As the learnt feature captures the relation between audio signals and their corresponding machine IDs, it offers a better ability to identify the sound from different IDs and enhance the relation between sounds from the same ID.

Supposing $\bm{X} = [\bm{x}_1, \cdots, \bm{x}_i, \cdots, \bm{x}_N]$ is a set of input audio signals that includes $N$ machine sounds. We select the $i$-th machine sound $\bm{x}_i$ ($1 \le i \le N$) as the anchor, and build the contrast with the remaining $(N - 1)$ audio signals. The machine ID label of $\bm{x}_i$, defined as $l_i$, and its audio embedding $\bm{z}_i \in \mathbb{R}^D$ can be extracted via the backbone and the projector modules, as shown in Fig. \ref{fig:CLPMetaSystem}. 
For the remaining $(N - 1)$ audio signals $\bm{x}_j$ ($1 \le j \le N, j \ne i$), we can obtain their corresponding ID label $l_j$ and audio embedding $\bm{z}_j$ in the same way. For contrastive learning, we can use the cosine similarity score defined in terms of the audio embeddings $\bm{z}_i$ and $\bm{z}_j$ as
\begin{equation}
\label{eq:cos}
    s_{i, j} = \frac{\bm{z}_i^\top * \bm{z}_j}{\| \bm{z}_i \|_2 \| \bm{z}_j \|_2}
\end{equation}
where $*$ denotes matrix multiplication, $\top$ represents transposition operation, and $\| \cdot \|_2$ is the $l_2$-norm function.

To capture the relation of the audio embeddings from the same ID and distinguish audio embeddings from different IDs, the cosine similarity score of audio embeddings from the same ID is expected to be maximized, whereas the cosine similarity score of the embeddings from different IDs to be minimized. Therefore, following \cite{NEURIPS2020_d89a66c7}, our CL-Meta loss for audio feature learning can be defined as 
\begin{equation}
\label{eq:CLloss}
    {L}_\text{CL-Meta} \!=\! - \frac{1}{N}\! \sum_{i=1}^{N}\! \frac{1}{|K(i)|}\! \sum_{k \in K(i)} \log{\frac{{\exp (s_{i, k} / \tau)}}{\sum_{j \ne i}^N{\exp (s_{i, j} / \tau)}}}
\end{equation}
where $\tau$ is the temperature scalar to scale the cosine similarity scores, which is used to enlarge the distance between audio embeddings from different IDs.  $K(i) \!=\! \{k | 1 \!\le\! k \!\le\! N, \text{and } k \!\ne\! i, l_i \!=\! l_k\}$ denotes the set of indexes that have the same ID as audio index $i$. $k$ is an index from $K(i)$, and $|K(i)|$ represents the number of indexes in $K(i)$.

With the contrastive learning loss Eq.~\eqref{eq:CLloss}, we can obtain a more effective audio representation to enhance the relation between sound from the same ID and the difference between different machine sounds. We can then use the learnt model parameters for the initialization of the model in the fine-tuning stage.

\subsection{Self-supervised Classification Based Fine-tuning}
\label{subsec:IDclassification}
With the pretrained audio representation, we then fine-tune our model by a self-supervised ID classifier with ArcFace loss \cite{deng2019arcface}  to further enhance the distinguishing ability of the learnt audio representation. 

Note that, a different model structure is applied in this stage as illustrated in Fig.~\ref{fig:CLPMetaSystem} (a) and (b), where a simple classifier is introduced after the projector to learn the latent feature $\bm{h}_i$ from the audio embedding $\bm{z}_i$ for ID prediction. Then, following \cite{liu2022anomalous}, we employ the self-supervised classification loss, i.e., ArcFace loss \cite{deng2019arcface} for the model fine-tuning, which can further improve the ability to distinguish the audio features from different IDs. The ArcFace loss is calculated as 
\begin{equation}
    {L}_\text{ArcFace} = \text{ArcFace}(\bm{h}_i, l_i).
\end{equation}
For the anomalous sound detection, we use the proposed CLP-SCF method to predict the ID of an estimated machine sound, and calculate the negative log probability of the estimated machine sound and its corresponding ID as the anomaly score for anomalous sound detection. That is, a normal sound is less likely to be predicted as a non-corresponding ID. Therefore, in the inference stage, if the predicted ID differs from the actual ID, it will be considered an anomalous sound.
\section{Experiments and Results}
\label{sec:experiment}
\subsection{Experimental Setup}
\textbf{\textit{Dataset}} 
Following \cite{liu2022anomalous}, our CLP-SCF method is evaluated on the DCASE 2020 Challenge Task2 development and additional datasets, which include four machine types (i.e., Fan, Pump, Slider and Valve) from the MIMII dataset \cite{Purohit_DCASE2019_01} and two machine types (i.e., ToyCar and ToyConveyor) from the ToyADMOS dataset \cite{Koizumi_WASPAA2019_01}. Each machine type has seven different machines, except for ToyConveyor, which only has six different machines. Therefore, we have audio signals from 41 different machines (41 ID labels), where each audio signal is around 10 seconds. The training data of the development dataset and the additional dataset are combined as the training set, and our model is trained for all machine IDs. The normal and anomalous sound from the test data of the development dataset is adopted for model evaluation. 

Note that in our experiments, the DCASE 2022 Challenge Task2 dataset \cite{Dohi_arXiv2022_02} was not used since it is designed to investigate domain shift, where the distribution of audio features of machine sounds may change from the known source domain to the unknown target domain. This is out of the scope of our work here, as we address the audio representation of sounds under known status. Therefore, we use DCASE 2020 dataset instead in our experiments.
\begin{table*}[!ht]
    \setlength{\abovecaptionskip}{0cm}
    \centering
    \caption{Performance comparison in terms of AUC (\%) and pAUC (\%) on the test data of the development dataset.}
    \resizebox{0.93\textwidth}{!}{
    \begin{tabular}{ccccccccccccccc}
        \toprule
        \multirow{2}{*}{Methods} & \multicolumn{2}{c}{Fan} & \multicolumn{2}{c}{Pump} & \multicolumn{2}{c}{Slider} & \multicolumn{2}{c}{Valve} & \multicolumn{2}{c}{ToyCar} & \multicolumn{2}{c}{ToyConveyor} & \multicolumn{2}{c}{Average} \\
        \cmidrule(lr){2-3} \cmidrule(lr){4-5} \cmidrule(lr){6-7} \cmidrule(lr){8-9} \cmidrule(lr){10-11} \cmidrule(lr){12-13} \cmidrule(lr){14-15} 
         & AUC & pAUC & AUC & pAUC & AUC & pAUC & AUC & pAUC & AUC & pAUC & AUC & pAUC & AUC & pAUC \\
        \midrule
        IDNN \cite{suefusa2020anomalous} & 67.71 & 52.90 & 73.76 & 61.07 & 86.45 & 67.58 & 84.09 & 64.94 & 78.69 & 69.22 & 71.07 & 59.70 & 76.96 & 62.57 \\
        MobileNetV2 \cite{Giri2020a} & 80.19 & 74.40 & 82.53 & 76.50 & 95.27 & 85.22 & 88.65 & 87.98 & 87.66 & 85.92 & 69.71 & 56.43 & 84.34 & 77.74 \\
        Glow\_Aff \cite{dohi2021flow} & 74.90 & 65.30 & 83.40 & 73.80 & 94.60 & 82.80 & 91.40 & 75.00 & 92.20 & 84.10 & 71.50 & 59.00 & 85.20 & 73.90 \\
        STgram-MFN (ArcFace) \cite{liu2022anomalous} & 94.04 & 88.97 & 91.94 & 81.75 & 99.55 & 97.61 & 99.64 & 98.44 & 94.44 & 87.68 & 74.57 & \textbf{63.60} & 92.36 & 86.34 \\
        AADCL \cite{hojjati2022self} & 85.27 & 68.93 & 86.75 & 70.85 & 77.74 & 61.62 & 68.62 & 55.03 & 88.79 & 75.95 & 71.26 & 57.40 & 79.74 & 64.96 \\
        \textbf{CLP-SCF} & \textbf{96.98} & \textbf{93.23} & \textbf{94.97} & \textbf{87.39} & \textbf{99.57} & \textbf{97.73} & \textbf{99.89} & \textbf{99.51} & \textbf{95.85} & \textbf{90.19} & \textbf{75.21} & 62.79 & \textbf{93.75} & \textbf{88.48} \\
        \bottomrule
    \end{tabular}
    }
    \label{tab:overall_performance}
    \vspace{-5mm}
\end{table*}

\noindent \textbf{\textit{Implementation Details}} 
In the pretraining stage, we randomly select 6 machine sounds from each ID to construct the set of input audio signals with the batch size of 246 ($41 \times 6$), which is used to build the contrast for the signals from all ID labels. Adam optimizer \cite{kingma2014adam} with a learning rate of 0.0005 is used for model optimization, and the model is pretrained with 100 epochs. The temperature score $\tau$ in Eq. \eqref{eq:CLloss} is empirically selected as 0.05 following \cite{NEURIPS2020_d89a66c7}.

In the fine-tuning stage, the batch size is converted to 128, the learning rate is set as 0.0001, and our model is fine-tuned with 300 epochs. For the self-supervised classification, the margin and scale hyper-parameters of the ArcFace loss are set as 1.0 and 30, respectively. 
Note that,  the cosine annealing strategy is adopted as the learning rate decay schedule in both stages \cite{loshchilov2016sgdr}. 

\noindent \textbf{\textit{Performance Metrics}} 
Following \cite{Koizumi_DCASE2020_01, liu2022anomalous, suefusa2020anomalous, Giri2020a, dohi2021flow}, we employ the area under the receiver operating characteristic curve (AUC) and the partial-AUC (pAUC) for performance evaluation. Here, pAUC denotes the AUC value over a low false-positive rate range $[0, p]$, where $p$ is set as 0.1 following \cite{Koizumi_DCASE2020_01, liu2022anomalous}. Meanwhile, minimum AUC (mAUC) is also adopted for detection stability evaluation, which reflects the worst detection performance of the machines from the same machine type \cite{dohi2021flow,liu2022anomalous}. 

\subsection{Performance Comparison}
\label{subsec:performance}
To show the performance of the proposed CLP-SCF, we compare our method with the state-of-the-art methods on DCASE 2020 Task2 dataset, including IDNN \cite{suefusa2020anomalous}, MobileNetV2 \cite{Giri2020a},  Glow\_Aff \cite{dohi2021flow}, STgram-MFN (ArcFace) \cite{liu2022anomalous} and AADCL \cite{hojjati2022self}. Here, IDNN is the AE-based method without machine information, and MobileNetV2, Glow\_Aff, and STgram-MFN (ArcFace) are the state-of-the-art self-supervised classification methods that also adopt machine ID for anomaly detection. The AADCL is the method using contrastive learning to learn audio representation via data augmentation, without exploring the relation between machine sound and its corresponding metadata. The results are shown in Table~\ref{tab:overall_performance}.

Our proposed method can achieve the best performance in terms of average AUC and average pAUC on all machine types, which provides 1.39\% and 2.14\% improvements in terms of average AUC and average pAUC, respectively, over the second best method, i.e., STgram-MFN (ArcFace), which is the backbone of our method. Except for the pAUC performance on ToyConveyor, the proposed method offers better detection performance in terms of both AUC and pAUC for all machine types, than the state-of-the-art self-supervised classification methods that also adopt machine IDs as self-supervision labels, and the contrastive learning based ASD method, i.e., AADCL. 
\begin{table}[t]
    \vspace{-3mm}
    \centering
    \caption{Performance comparison in terms of mAUC (\%).}
    \resizebox{0.9\columnwidth}{!}
    {
    \begin{tabular}{ccc}
        \toprule
        Methods & STgram-MFN (ArcFace) \cite{liu2022anomalous} & \textbf{CLP-SCF} \  \\
        \midrule
        Fan & 81.39 & \textbf{88.27}  \\
        Pump & 83.48 & \textbf{87.27}  \\
        Slider & 98.22 & \textbf{98.28}  \\
        Valve & 98.83 & \textbf{99.58}  \\
        ToyCar & 83.07 & \textbf{86.87}  \\
        ToyConveyor & 64.16 & \textbf{65.46}  \\
        \midrule
        Average & 84.86 & \textbf{87.62}  \\
        \bottomrule
    \end{tabular}
    }
    \label{tab:mAUC}
    \vspace{-5mm}
\end{table}
\subsection{Detection Stability}
\label{subsec:stability}
Table~\ref{tab:mAUC} presents the mAUC performance of our CLP-SCF, as compared with that of STgram-MFN (ArcFace) \cite{liu2022anomalous}. The study in  \cite{liu2022anomalous} significantly improved the detection stability and performance via its proposed spectral-temporal fusion feature (STgram). From Table~\ref{tab:mAUC}, we can see that our proposed method can further improve detection stability with significant improvement in mAUC for all machine types, especially for the machine types of Fan, Pump, and ToyCar. The results from Tables~\ref{tab:overall_performance} and \ref{tab:mAUC} verify the effectiveness of the proposed method for improving detection performance and stability.

\begin{figure}[!t]
	\centering
	\subfloat[\centering{STgram-MFN (ArcFace)  \cite{liu2022anomalous}}]{
	    \label{subfig:stgram}
        \includegraphics[width=0.93\linewidth]{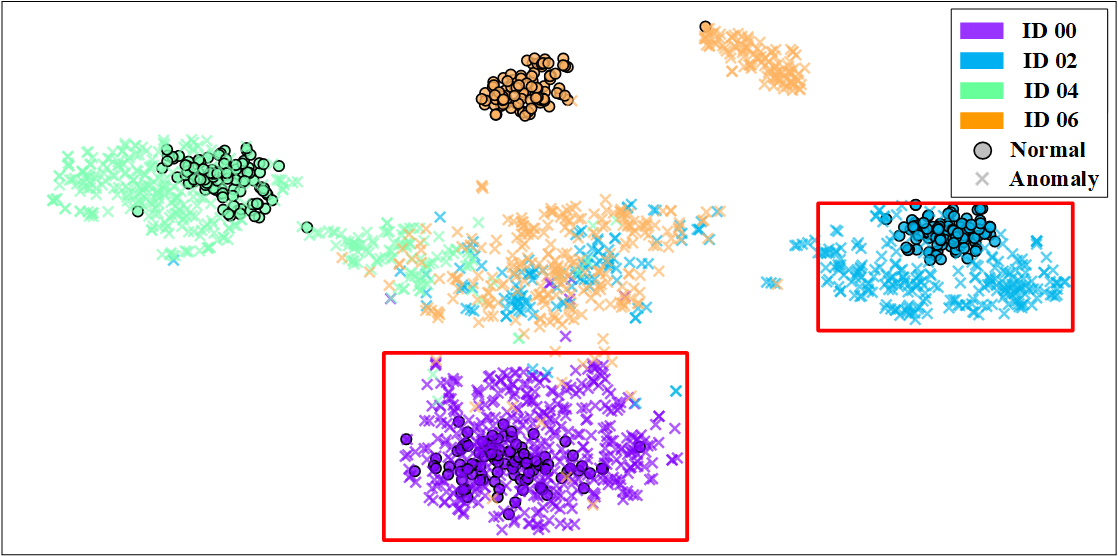}
	    }
	\vspace{-3mm}
	\quad
	\subfloat[\centering{Proposed CLP-SCF}]{
	    \label{subfig:stgram_CLPMeta}
        \includegraphics[width=0.93\linewidth]{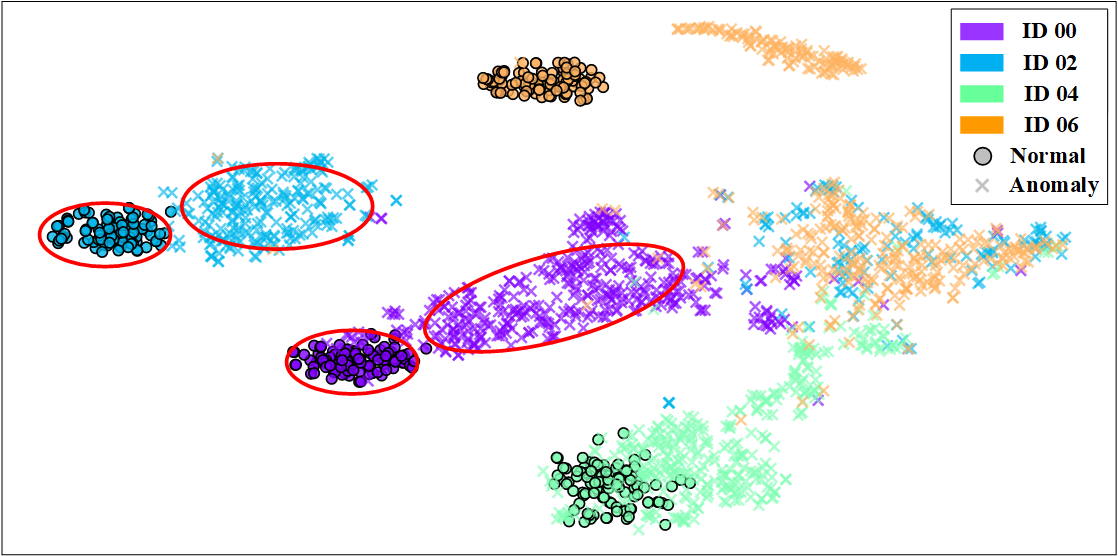}
	    }
	\vspace{-3mm}
    \caption{The t-SNE visualization of STgram-MFN (ArcFace) \cite{liu2022anomalous} and the proposed CLP-SCF for the machine type Fan. (a) denotes the latent feature distribution obtained using the STgram-MFN (ArcFace) method. (b) denotes the latent feature distribution obtained using the proposed CLP-SCF method. The symbol ``$\bullet$" and ``$\times$" denote normal and anomalous sound classes, respectively. The normal and anomalous latent feature distributions for ``ID 00" and ``ID 02" are highlighted by the red contours.}
	\label{fig:tsne}
	\vspace{-5mm}
\end{figure}

To illustrate the effect of the learnt audio feature representation, Fig.~\ref{fig:tsne} shows the t-distributed stochastic neighbour embedding (t-SNE) cluster visualization of the latent features of these two methods, where we can see that our method shows better distinguishing ability. For example, compared to STgram-MFN (ArcFace), the proposed method significantly reduces the overlapping between the normal and anomalous latent features of ``ID 00'' and ``ID 02'' in Fig.~\ref{fig:tsne}. The result further demonstrates the effectiveness of the proposed method. 
\section{Conclusion}
\label{sec:conclusion}
In this paper, we have studied the relation between the metadata and the machine sound in audio representation for anomalous sound detection. We have presented a two-stage method to improve the quality of the audio representation, which consists of model pretraining using the metadata-based contrastive learning in the first stage, and model fine-tuning using the self-supervised ID classification in the second stage. Experiments show that the proposed method achieves better detection performance than the state-of-the-art methods. 


\bibliographystyle{IEEEtran}
\bibliography{strings}

\end{document}